\documentclass[aps,prb,twocolumn,groupedaddress,showpacs]{revtex4}
\usepackage{amsmath}
\usepackage{amssymb}
\usepackage{graphicx}

\begin{document}

%\preprint{}


\title{Curved One-Dimensional Wire as a Spin Rotator}


\author{Maxim P. Trushin and  Alexander L. Chudnovskiy}
%\thanks{}
\affiliation{1. Institut f\"ur Theoretische Physik, Universit\"at Hamburg,
Jungiusstr 9, D-20355 Hamburg, Germany}


\date{\today}

\begin{abstract}
We propose a semiconductor structure that can rotate the electron spin without using
ferromagnetic contacts, tunneling barriers, external radiation etc.
The structure consists of a strongly curved one-dimensional ballistic wire
with intrinsic spin-orbit interactions of Rashba type. Our calculations and analytical
formulae show that the proposed device can redistribute the current densities
between the two spin-split modes without backscattering and, thus, serve as a
reflectionless and high-speed spin switcher.
Using parameters relevant for InAs we investigate the projection of current density 
spin polarization on the spin-quantization axis as a function of the 
Rashba constant, external magnetic field, and radius of the wire's curvature.
\end{abstract}


%\keywords{spintronics, Rashba spin-orbit interactions}

\maketitle
%\showpacs

\section{Introduction}

In the past few years
the idea to use electron spin in mesoscopic semiconductor devices has generated a lot of interest.
Datta and Das \cite{APL1990datta} describe how Rashba effect \cite{JphC1984bychkov}
(with the assistance of spin-filtering contacts) can be used to modulate the current.  
The basic idea is that the spin precession can be controlled via Rashba spin-orbit coupling
associated with the interfacial electric field present in the quantum well that
contains a two-dimensional electron gas. One of the most
promising materials for this purpose is the InAs semiconductors,
where the tuning of the Rashba coupling by an external gate voltage was recently 
achieved by Grundler \cite{PRL2000grundler} and Matsuyama et al. \cite{PRB2000matsuyama}

\begin{figure}
\includegraphics{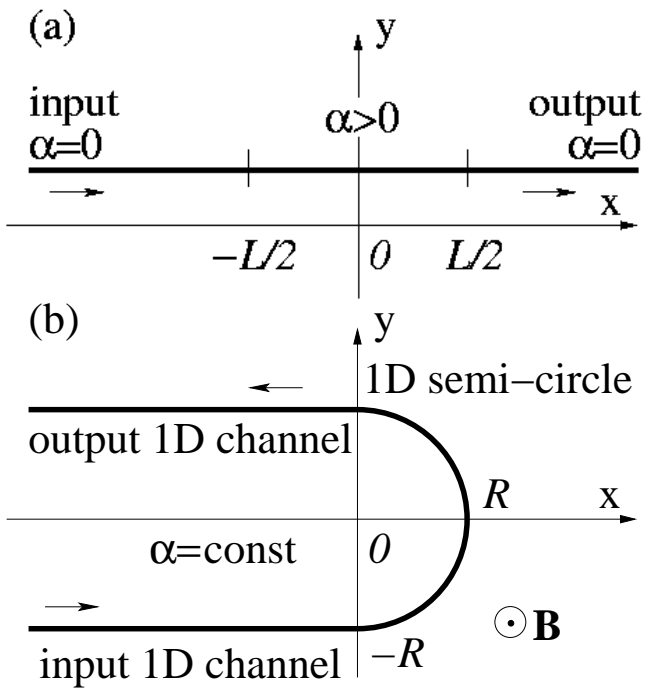}
\caption{\label{fig1} (a) Schematic of the ``conventional'' spin-rotator.
In the central region of the wire the spin-orbit interaction is finite, whereas
in the input and output channels the Rashba effect vanishes.
(b) Schematic of the reflectionless and high-speed spin-rotator. 
The quantum wire is made of just one material so, that
the Rashba constant in the curved part is the same as in the input and output channels.}
\end{figure}

The schematic of the ``conventional'' spin-rotator based on Rashba effect is
depicted in Fig.~\ref{fig1}a. 
The straight quantum wire (or just a two-dimensional stripe)
is divided into three regions.
In the middle region of the length $L$ the spin-orbit interactions are finite 
at the semiconductor interface, whereas in the input and output channels
the spin-orbit coupling is set to zero. 
In other words, the semiconductor interface at which the Rashba effect occurs does not extend into
the regions connected to the ferromagnetic source and drain \cite{PRB2001mireles}.
The angle of the spin rotation depends explicitely on the length of the stripe 
between the input and output contacts, namely
\begin{equation}
\Delta\vartheta= \frac{2 m^* \alpha L}{\hbar^2},
\end{equation}
where $m^*$ is the effective electron mass, $\alpha$ is the Rashba constant.

Let us estimate the spin-switching speed of this device in the ballistic transport regime.
In other words, we need the minimal time necessary to rotate the spin
for the angle of $\Delta\vartheta=\pi$. We take the parameters relevant for InAs, i. e.
Rashba constant is $\alpha=2\cdot 10^{-11}$eVm, whereas 
the effective electron mass is $m^*=0.033 m_e$. 
Then, the characteristic length of the ``active'' region necessary to rotate the electron spin
to its opposite direction is equal to $\sim 10^{-5}$cm.
Assuming the characteristic velocity $5\cdot 10^7$cm/s
we have, that the cycle time is $0.2$ps that corresponds to $5$ THz.
Thus, the hypothetical spintronic transistor might be one thousand times faster
than conventional one.

In spite of the impressive advantages, the abovementioned schema involves propagation
of electrons across borders separating the media with different spin-orbit
coupling strength. A reflection on the border 
is thus a necessary complement that diminishes the total current through
the device and even might compromise the feasibility of the proposal.
In this paper, we propose a scheme of the {\em reflectionless} spin-rotator made of material
with Rashba spin-orbit interaction such as InAs. 
We consider a {\em curved} wire consisting of a semicircle with radius $R$ attached
to the infinite straight one-dimensional channels, as shown in Fig.~\ref{fig1}b.
The channels are made of the same material as the semicircle itself, thus,
the electron backscattering is negligible.
Moreover, because of specific geometry of the system,
the speed of response can even exceed the one for the ``conventional'' spin-rotator
discussed above. The device is placed in a perpendicular magnetic field ${\mathbf B}$,
which can be used to control the spin-rotation (in addition to the gate-voltage).
Curved one-dimensional quantum channels in InAs
\cite{PRB2002yang} are expected to be used for the experimental check of the present proposal.
The spin polarized electrons necessary for such experiments can be generated in InAs by circularly polarized light \cite{PRL2001ganichev,PRB2003ganichev}. Note, that the recombination of spin polarized charged carriers
results in the emission of circularly polarized light. It is possible, therefore,
to use the optical methods for the detection of the spin-polarization as well.

On the face of it, the device depicted in Fig.~\ref{fig1} is similar to the one investigated
by Bulgakov and Sadreev \cite{PRB2002bulgakov}. However, there is an essential difference in 
approaches used here and in Ref.~\cite{PRB2002bulgakov}. In that work, the authors 
assume {\em a priory} the adiabatic regime: the radius of the curvature is 
so large that the electrons do not feel the junction between the curved part of the wire
and input/output channels. In contrast, we start from the very general solution of
Schr\"odinger equation for the {\em whole} system 
(i. e. input channel --- semi-circle --- output channel)
and find that though the electron backscattering is still negligible, the redistribution
between current densities with opposite spin indices can occur at certain
comparatively small radii of curvature (that is forwardscattering in some sense).

In order to describe the degree of the current density redistribution 
between the modes with opposite spin indices we introduce the following
quantity
\begin{equation}
P=\frac{j^+ - j^-}{j^+ + j^-},
\label{polarization}
\end{equation}
where $j^{\pm}$ denote the current densities \cite{landau1958} with a given spin orientation, and
``$\pm$'' are the spin indices.
It seems essential to emphasize that the quantity $P$ is controllable experimentally since
the currents $j^+$ and $j^-$ can be generated and detected independently by means
of absorption of two circularly polarized light beams with opposite helicity \cite{PRL2001ganichev,PRB2003ganichev}.

If one prefers to control $P$ by means of magnetized contacts then the situation is a bit more
complicated. Indeed, the quantity $P$
has the meaning of projection of the current density spin-polarization
on the spin-quantization axis whose orientation is determined by the relation
between the external magnetic field ${\mathbf B}$ and in-plane Zeeman-like
magnetic field $B_\mathrm{in}$ generated by the Rashba spin-orbit interactions
\begin{equation}
B_\mathrm{in}=\frac{2\alpha k_0}{g^* \mu_B}.
\end{equation}
Here $k_0$ is the characteristic Fermi wave vector, whereas
$g^*$ and  $\mu_B$ being g-factor and Bohr magneton respectively.
The field ${\mathrm B}_\mathrm{in}$ is orthogonal to the direction of the electron motion,
therefore the spin-quantization axis lies in the $yz$ plane (for the input and output channels).
The angle $\gamma_0$ between the $z$ axis and the spin-quantization one
can be found from the simple trigonometric formula
\begin{equation}
\tan\gamma_0=\frac{B_\mathrm{in}}{|{\mathbf B}|}.
\end{equation}
If the external magnetic field $|{\mathbf B}|=B_z$ is much larger than 
the in-plane one, then the spin-quantization axis coincides with $z$ axis.
In contrast, if the external magnetic field is absent then
the spin-quantization axis is orthogonal to the direction of the electron motion
at each point of its trajectory. 
In the following, we call the quantity $P$ defined by (\ref{polarization})
just spin-polarization.

In the next sections we outline our calculation of the current density
spin-polarization and discuss the results obtained.

\section{General solution}

We calculate single particle spin-split states for the system shown in Fig.~\ref{fig1}b.
To this end, we divide the wire in three parts: input channel, semi-circle (curved part of the
quantum wire) and output channel.
We use the cartesian coordinates to describe the input and output channels 
(the region $x < 0$ in Fig.~\ref{fig1}b) and the polar coordinates
for the description of the curved part of the wire (the semi-circle).
The Hamiltonians describing the propagation of an electron in the input/output wires read
\begin{equation}
\label{1Dwire_hamiltonian}
H_\mathrm{wire}=\left( \begin{array}{cc}
\frac{\hbar^2}{2m^*}\hat{k}_x^2 + \varepsilon_Z &
i\alpha\, \hat{k}_x \\
-i\alpha\, \hat{k}_x  &
\frac{\hbar^2}{2m^*}\hat{k}_x^2 - \varepsilon_Z \end{array} \right)
\end{equation}
whereas the propagation through the semi-circle of radius $R$ is governed by the Hamiltonian
\cite{PRB2002meijer}
\begin{equation}
\label{loop_hamiltonian}
H_\mathrm{curv}=\left( \begin{array}{cc}
\varepsilon_0\,\hat{q}_{\varphi}^2 + \varepsilon_Z &
\alpha\,{\mathrm e}^{-i\varphi}\left(\hat{q}_\varphi - \frac{1}{2}\right)/R \\
\alpha\,{\mathrm e}^{i\varphi}\left(\hat{q}_\varphi + \frac{1}{2}\right)/R  &
\varepsilon_0\,\hat{q}_{\varphi}^2 - \varepsilon_Z \end{array} \right).
\end{equation}
Here $\hat{k}_x=-i\frac{\partial}{\partial x}-\frac{\Phi}{\Phi_0}\frac{1}{R}$,
$\hat{q}_\varphi=-i\frac{\partial}{\partial\varphi}-\frac{\Phi}{\Phi_0}$
are the momentum and the angular momentum operators respectively,
$\Phi=\pi\,R^2 B_z$ is a magnetic flux through the area of a ring of radius $R$,
$\Phi_0=2\pi\,\hbar c/e$ is the flux quantum,
$\varepsilon_0=\hbar^2/(2m^*R^2)$ is the size confinement energy,
$\varepsilon_Z=g^*\mu_B B_z/2$ is the Zeeman term.
We adopt the vector potential $\mathbf{A}$ to be tangential to the direction of the current.
Thus, in the semi-circle we choose $\mathbf{A}(x,y)=\frac{1}{2}B_z\left(x\,\mathbf{j}-y\,\mathbf{i}\right)$,
or, in cylindrical coordinates, $A_\varphi (\varphi)=\Phi/2\pi R$, whereas
the vector potential in the input and output channels is determined by the continuity condition 
at the junction point with the curved part of the wire 
($x=0$, $y=\pm R$); hence we have $A_x=\Phi/2\pi R$.

We denote the wave functions for each part as $\Psi_\mathrm{curv}^\pm(\varphi)$ for the semi-circle,
$\Psi_\mathrm{in}^\pm(x)$ and $\Psi_\mathrm{out}^\pm(x)$ for the input and the output
channels respectively. In order to find the wave function of the whole system,
we impose boundary conditions that warrant the continuity of the wave function and its first derivative on the boundaries between the parts of the wire
\begin{equation}
\label{conditions}
\left\{\begin{array}{l}
\left(\Psi_\mathrm{in}^+ + \Psi_\mathrm{in}^-\right)|_{x=0}
=\left(\Psi_\mathrm{curv}^+ + \Psi_\mathrm{curv}^-\right)|_{\varphi=-\pi/2}, \\
\left(\Psi_\mathrm{curv}^+ + \Psi_\mathrm{curv}^-\right)|_{\varphi=\pi/2}=
\left(\Psi_\mathrm{out}^+ + \Psi_\mathrm{out}^-\right)|_{x=0}, \\
\left(\nabla\Psi_\mathrm{in}^+ + \nabla\Psi_\mathrm{in}^-\right)|_{x=0}=
\left(\nabla\Psi_\mathrm{curv}^+ + \nabla\Psi_\mathrm{curv}^-\right)|_{\varphi=-\pi/2}, \\
\left(\nabla\Psi_\mathrm{curv}^+ + \nabla\Psi_\mathrm{curv}^-\right)|_{\varphi=\pi/2}=
\left(\nabla\Psi_\mathrm{out}^+ + \nabla\Psi_\mathrm{out}^-\right)|_{x=0}.
\end{array}\right.
\end{equation}
Solutions of  Schr\"odinger equations for Hamiltonians (\ref{1Dwire_hamiltonian}), (\ref{loop_hamiltonian})
give us the desired spinor wave functions for the input, output and curved parts of the system.
For the input channel we have
\begin{equation}
\label{input_psi+}
\Psi^+_\mathrm{in}(x)={\mathrm e}^{\frac{i \Phi }{\Phi_0 R}x}
\left(\begin{array}{c}
\cos\gamma^+\left(A_0^+{\mathrm e}^{i\theta^+ + i k^+ x} + A^+ {\mathrm e}^{-i k^+ x}\right) \\
-i\sin\gamma^+\left(A_0^+{\mathrm e}^{i\theta^+ + i k^+ x} - A^+ {\mathrm e}^{-i k^+ x}\right) \end{array} \right),
\end{equation}
\begin{equation}
\label{input_psi-}
\Psi^-_\mathrm{in}(x)={\mathrm e}^{\frac{i \Phi }{\Phi_0 R}x}
\left( \begin{array}{c}
-i\sin\gamma^-\left(A_0^-{\mathrm e}^{i\theta^- + i k^- x} - A^- {\mathrm e}^{- i k^- x}\right) \\
\cos\gamma^-\left(A_0^-{\mathrm e}^{i\theta^- + i k^- x} + A^- {\mathrm e}^{-i k^- x}\right) \end{array} \right),
\end{equation}
where
\begin{equation}
\tan\gamma^\pm= -\frac{{\varepsilon_Z}}{{k^{\pm}}\,{\alpha}}  +
\sqrt{1 + \left(\frac{{\varepsilon_Z}}{{k^{\pm}}\,{\alpha}}\right)^2}.
\end{equation}
Here $\theta^\pm$ are the initial phases,
and $k^{\pm}$ are the Fermi wave vectors that satisfy the dispersion relations
\begin{equation}
\label{spectrum_wire}
E_F=\frac{\hbar^2 {k^\pm}^2}{2m^*}\pm\sqrt{\alpha^2 {k^\pm}^2+\varepsilon_Z^2},
\end{equation}
where $E_F$ is the Fermi energy.
In the case of zero magnetic field ($\varepsilon_Z=0$), the Fermi momenta $k^\pm$ 
take the simple form
\begin{equation}
\label{Fermi_k_noB}
k^{\pm}=\mp\frac{m^*\,\alpha}{\hbar^2}+k_0,
\end{equation}
where $k_0=\sqrt{\left(m^*\,\alpha/\hbar^2\right)^2 + 2m^*\,E_F/\hbar^2}$.
The coefficients $A^\pm$ are the reflection amplitudes that have to be found by imposing
the boundary conditions (\ref{conditions}), whereas $A_0^\pm$ are 
the incident ones.

For the output channel the reflection amplitudes are assumed to be zero,
and the corresponding spinors read
\begin{equation}
\label{out_psi+}
\Psi^+_\mathrm{out}(x)=\left(\begin{array}{l}
D^+\cos\gamma^+ {\mathrm e}^{i(k^+ + \frac{\Phi}{\Phi_0 R})x} \\
iD^+\sin\gamma^+ {\mathrm e}^{i(k^+ + \frac{\Phi}{\Phi_0 R})x} \end{array} \right),
\end{equation}
\begin{equation}
\label{out_psi-}
\Psi^-_\mathrm{out}(x)=\left( \begin{array}{l}
iD^-\sin\gamma^-  {\mathrm e}^{i(k^- + \frac{\Phi}{\Phi_0 R})x} \\
D^-\cos\gamma^- {\mathrm e}^{i(k^- + \frac{\Phi}{\Phi_0 R})x} \end{array} \right).
\end{equation}
Here $D^\pm$ are the transmission amplitudes.
We have changed the sign of $\gamma^\pm$ for
the output wire since the electron motion changes its direction to the opposite one.

The eigenfunctions of the Hamiltonian (\ref{loop_hamiltonian}) have a view
\begin{widetext}
\begin{equation}
\label{loop_psi+}
\Psi^+_\mathrm{curv}(\varphi)={\mathrm e}^{i \frac{\Phi}{\Phi_0} \varphi}
\left(
\begin{array}{l}
B^+ \cos\alpha^+ {\mathrm e}^{i(q^+_R - \frac{1}{2})\varphi}
+C^+\cos\beta^+ {\mathrm e}^{-i(\frac{1}{2} + q^+_L)\varphi}  \\
B^+ \sin\alpha^+ {\mathrm e}^{i(\frac{1}{2} + q^+_R)\varphi}
-C^+\sin\beta^+ {\mathrm e}^{-i(q^+_L - \frac{1}{2})\varphi}  \end{array}
\right),
\end{equation}
\begin{equation}
\label{loop_psi-}
\Psi^-_\mathrm{curv}(\varphi)={\mathrm e}^{i \frac{\Phi}{\Phi_0} \varphi}
\left(\begin{array}{l}
- B^- \sin\alpha^- {\mathrm e}^{i(q^-_R - \frac{1}{2})\varphi}
+C^-\sin\beta^- {\mathrm e}^{-i(\frac{1}{2} + q^-_L)\varphi}  \\
B^- \cos\alpha^- {\mathrm e}^{i(\frac{1}{2} + q^-_R)\varphi}
+C^-\cos\beta^- {\mathrm e}^{-i(q^-_L - \frac{1}{2})\varphi}  \end{array} \right),
\end{equation}
\end{widetext}
where
\begin{equation}
\tan\alpha^{\pm}= \frac{{\varepsilon_0 q^\pm_R-\varepsilon_Z}}{{q^{\pm}_R}\,{\alpha}/R}  +
\sqrt{1 + \left(\frac{\varepsilon_Z-\varepsilon_0 q^\pm_R}{q^{\pm}_R \,\alpha /R}\right)^2},
\end{equation}
\begin{equation}
\tan\beta^{\pm}= - \frac{{\varepsilon_0 q^{\pm}_L + \varepsilon_Z}}{{q^{\pm}_L}\,{\alpha}/R}  +
\sqrt{1 + \left(\frac{\varepsilon_Z+\varepsilon_0 q^\pm_L}{{q^{\pm}_L}\,{\alpha}/R}\right)^2},
\end{equation}
and $q^{\pm}_{R,L}$ are the Fermi angular momenta in the curved part of the wire that are found from the
conditions
\begin{equation}
\label{spectrum_loopR}
E_F = \frac{{\varepsilon_0}}{4} + \varepsilon_0{q_R^\pm}^2
\pm\sqrt{\left(\frac{{{q_R^\pm}}\,{{\alpha}}}{R}\right)^2 + {\left({q_R^\pm}\,{\varepsilon_0}
 -  {\varepsilon_Z} \right) }^2},
\end{equation}
\begin{equation}
\label{spectrum_loopL}
E_F = \frac{{\varepsilon_0}}{4} + \varepsilon_0{q_L^\pm}^2
\pm\sqrt{\left(\frac{{{q_L^\pm}}\,{{\alpha}}}{R}\right)^2 + {\left({q_L^\pm}\,{\varepsilon_0}
 +  {\varepsilon_Z} \right) }^2}.
\end{equation}
If the Zeeman effect is negligible, then the equations (\ref{spectrum_loopR}) and (\ref{spectrum_loopL}) 
allow the simple analytical solution with respect to $q^{\pm}_{R,L}$ 
\begin{equation}
\label{q_noB}
q^{\pm}/R=\mp\frac{m^*\,\alpha}{\hbar^2}\sqrt{1+\left(\frac{\hbar^2}{2\alpha\,m^* R}\right)^2}+k_0.
\end{equation}
Note, that the chirality index is omitted in (\ref{q_noB}), since $q^{\pm}_R=q^{\pm}_L$.

Imposing the boundary conditions (\ref{conditions}) on the wave functions (\ref{input_psi+}),
(\ref{input_psi-}), (\ref{out_psi+}) -- (\ref{loop_psi-}) we obtain a solution of the
Schr\"odinger equation for the whole system.
At this point it is pertinent to turn to the calculation of the current densities.
Using foregoing results one can easily find the input, reflected and
transmitted current densities. Each current density is given as a sum of its two spin-polarized parts $j=j^{+}+j^{-}$, where the components $j^\pm$ read
\begin{equation}
\label{j_in}
j_\mathrm{in}^\pm = \frac{\hbar}{m^*} |A_0^\pm|^2\left[k^{\pm}\pm\frac{\alpha m^*}{\hbar^2}\sin(2\gamma^\pm)\right],
\end{equation}
\begin{equation}
j_\mathrm{refl}^\pm =-\frac{\hbar}{m^*} |A^\pm|^2\left[k^{\pm}\pm\frac{\alpha m^*}
{\hbar^2}\sin(2\gamma^\pm)\right],
\end{equation}
\begin{equation}
j_\mathrm{out}^\pm =\frac{\hbar}{m^*} |D^\pm|^2\left[k^{\pm}\pm\frac{\alpha m^*}
{\hbar^2}\sin(2\gamma^\pm)\right].
\end{equation}
The transmission probability is defined as $T=j_\mathrm{out}/j_\mathrm{in}$,
and the reflection one as $R=j_\mathrm{refl}/j_\mathrm{in}$.
The general solution gives $T=1$ and $R=0$, which means that there is no particle backscattering.
However, there is  a current density redistribution between $j_\mathrm{out}^+$
and $j_\mathrm{out}^-$, which leads to the change of spin-polarization in the output current.
Some particular solutions of the system (\ref{conditions}) will be discussed in the next section.

\section{Discussion}

In this section we discuss two cases: (i) the input polarization is zero
$P_\mathrm{in}=0$, and (ii) $P_\mathrm{in}=1$.
Note, that in accordance with the definition (\ref{polarization})
the case $P_\mathrm{in}=0$ does {\em not} mean that the incident electron beam
is not spin-polarized at all, it rather means that the 
{\em projection} of the spin polarization on the spin-quantization axis is zero.

\subsection{$P_\mathrm{in}=0$}

In this subsection we assume, that $A_0^+=A_0^-$.
The general solution of the system (\ref{conditions}) is very cumbersome, thus,
we find the analytical expressions for the amplitudes 
$A^\pm$, $B^\pm$, $C^\pm$, $D^\pm$ in two limiting cases.
First, we assume zero external magnetic field ($\varepsilon_Z =0$, $\Phi=0$) and the adiabatic regime for
the spin precession $\hbar^2/(2m^*R\alpha)\ll 1$. In this case, one can adopt
$\gamma^\pm=\alpha^\pm=\beta^\pm=\pi/4$, and
the system of equations (\ref{conditions}) takes a simpler form (\ref{system_adiabatic}).
Note, that $k^\pm=q^\pm_R/R=q^\pm_L/R$ as long as $\hbar^2/(2m^*R\alpha)\ll 1$,
as it follows from the relations (\ref{Fermi_k_noB}) and (\ref{q_noB}).
Therefore, the solution of (\ref{system_adiabatic}) is rather trivial in this case:
$|A^+|^2=|A^-|^2=0$, and $|D^+|^2=|D^-|^2=1$. Then, the current densities read
\begin{equation}
j_\mathrm{out} =\frac{\hbar}{m^*}\sum\limits_\pm \left(k^{\pm}\pm\frac{\alpha m^*}{\hbar^2}\right),
\quad j_\mathrm{in} = j_\mathrm{out},
\end{equation}
and the polarization is
\begin{equation}
P_\mathrm{out}=\frac{k^{+}-k^{-}+2\alpha m^* /\hbar^2}{k^{+}+k^{-}}.
\end{equation}
Recall, that $k^{+}-k^{-}=-2\alpha m^* /\hbar^2$ at $B_z=0$. Thus, $P_\mathrm{out}=0$
for any values of $\alpha$.

\begin{figure}
\includegraphics{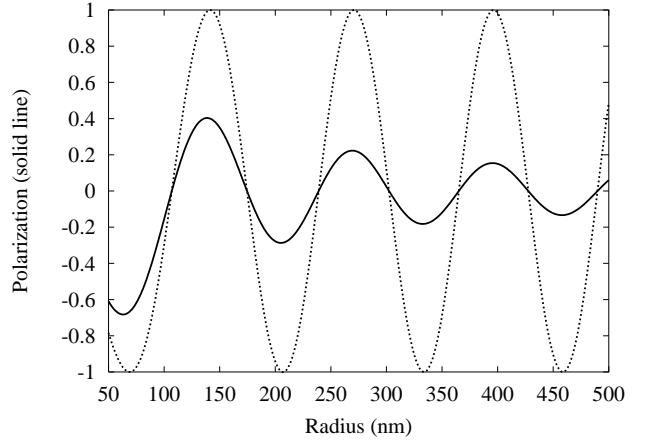}
\caption{\label{fig2} Polarization $P_\mathrm{out}$ (solid lines) and the interference factor
$\sin\left[\pi(q^-_R - q^+_R)\right]$
(dashed lines) versus radius of curvature at zero magnetic field.
The initial phases $\theta^\pm$ both are equal to zero, $P_\mathrm{in}=0$, and
the other parameters are taken relevant for InAs:
$\alpha=2\cdot 10^{-11}$eVm, $m^*=0.033 m_e$, $g^*=-12$, $E_F=30$meV.}
\end{figure}

%\begin{widetext}
\begin{figure*}
\includegraphics{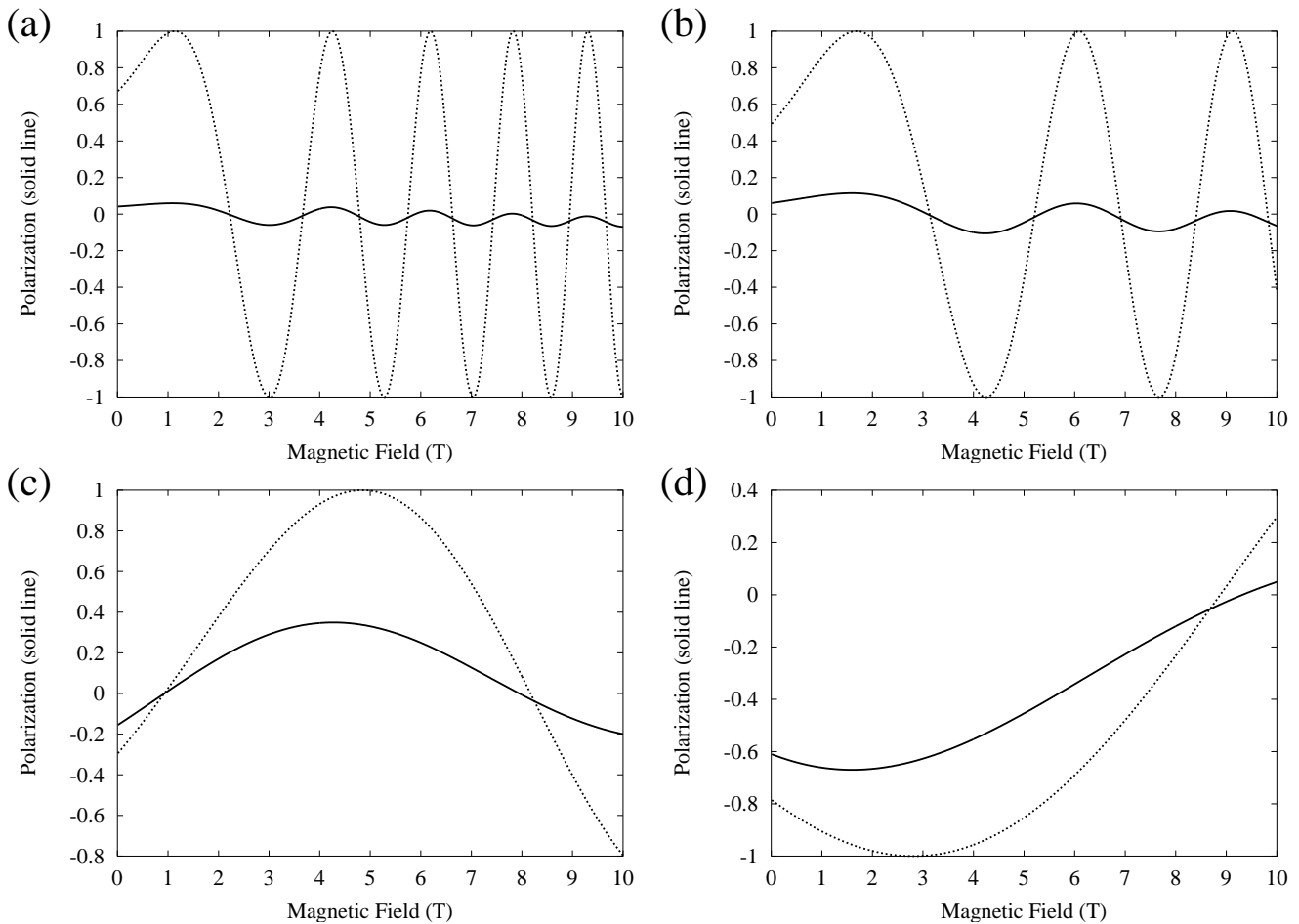}
\caption{\label{fig2a}
Polarization $P_\mathrm{out}$ (solid lines) and the interference factor
$\sin\left[\pi(q^-_R - q^+_R)\right]$
(dashed lines) versus external magnetic field at different radii of curvature:
(a) $R=10^{-4}$cm, (b) $R=5\cdot 10^{-5}$cm, (c) $R=10^{-5}$cm,
(d) $R=5\cdot 10^{-6}$cm.
The other parameters are the same as in Fig.~\ref{fig2}.}
\end{figure*}
%\end{widetext}

In contrast to that, if we assume the strongly non-adiabatic regime for
the spin precession so, that $\hbar^2/(2m^*R\alpha)\gg 1$, then 
at zero external magnetic field we still have 
$\gamma^\pm=\pi/4$, but $\alpha^\pm=\pi/2$ and $\beta^\pm=0$.
In this case, the system of equations (\ref{conditions}) takes the form
(\ref{system_non-adiabatic}), and the approximate solution reads
\begin{eqnarray}
\nonumber &&
A^+=0, \quad A^- = 0, \\
\nonumber &&
B^+ = \frac{1}{\sqrt{2}}\left({\mathrm e}^{i\theta^-} - i\,{\mathrm e}^{i\theta^+}\right)
{\mathrm e}^{\frac{i\,\pi}{2}\left(\frac{1}{2}+q^+_R\right)}, \\
\nonumber &&
B^- = -\frac{1}{\sqrt{2}}\left({\mathrm e}^{i\theta^+} - i\,{\mathrm e}^{i\theta^-}\right) {\mathrm e}^{\frac{i\,\pi}{2}\left(q^-_R - \frac{1}{2}\right)},\\
\nonumber &&
C^+=0, \quad C^- = 0, \\
\nonumber &&
D^+ = \frac{{\mathrm e}^{i\theta^-} + i\,{\mathrm e}^{i\theta^+}}{2i}{\mathrm e}^{i\,\pi\left(q^-_R - \frac{1}{2}\right)} +
\frac{{\mathrm e}^{i\theta^-} - i\,{\mathrm e}^{i\theta^+}}{2i}{\mathrm e}^{i\,\pi\left(q^+_R + \frac{1}{2}\right)}, \\
\nonumber &&
D^- = \frac{{\mathrm e}^{i\theta^-} - i\,{\mathrm e}^{i\theta^+}}{2}{\mathrm e}^{i\,\pi\left(q^+_R + \frac{1}{2}\right)} -
\frac{{\mathrm e}^{i\theta^-} + i\,{\mathrm e}^{i\theta^+}}{2}{\mathrm e}^{i\,\pi\left(q^-_R - \frac{1}{2}\right)}. \\
\label{non-ad-solution}
\end{eqnarray}
Then, $|D^\pm|^2=1\pm\cos\left(\theta^+ - \theta^-\right)\sin\left[\pi\,\left(q^-_R - q^+_R\right)\right]$,
and the spin components of the output current density read
\begin{equation}
j^\pm_\mathrm{out} = \frac{\hbar\, k_0}{m^*}\left\{1 \pm \cos\left(\theta^+ - \theta^-\right)
\sin\left[\pi\,\left(q^-_R - q^+_R\right)\right]\right\}.
\end{equation}
The spin components of the reflection current density 
$j^\pm_\mathrm{refl}$ are equal to zero.
Thus, $R=0$ and $T=1$, whereas the output polarization reads
\begin{equation}
\label{p_estimation}
P_\mathrm{out}=\cos\left(\theta^+ - \theta^-\right)\sin\left[\pi\,\left(q^-_R - q^+_R\right)\right].
\end{equation}

The relation (\ref{p_estimation}) shows, that in strongly curved one-dimensional wires the 
current density redistribution between the two spin-split modes is achievable.
The results of numerical calculations at $\theta^\pm=0$ are summarized in Figs.~\ref{fig2},~\ref{fig2a}.
The dependences $P_\mathrm{out}(R)$ and $P_\mathrm{out}(B_z)$ are given by solid curves.
The dotted lines correspond to the approximate expression (\ref{p_estimation}).
The strong correlation between the spin polarization and the interference factor is clearly visible.
Nevertheless, a few words of comment are necessary here.

First, the polarization is not zero at $B_z=0$. One can see that from the
Figs.~\ref{fig2},~\ref{fig2a} or directly from (\ref{p_estimation}).
Second, a plot of $P_\mathrm{out}$ as a function of $B_z$ ($R$ as well) yields an oscillating curve.
The oscillations have a natural explanation if one follows the evolution of the wave function
as a particle propagates through the wire. Namely, after entering into
the curved part of the wire, the
component of the input wave function $\Psi_\mathrm{in}^+$ propagates as a linear combination of the modes
$\Psi_\mathrm{curv}^+$ and $\Psi_\mathrm{curv}^-$ with the wave vectors $q^+_R$ and $q^-_R$ respectively
[see the approximate solution (\ref{non-ad-solution})].
The same is true for the propagation of the state $\Psi_\mathrm{in}^-$.
Due to the interference between the two propagated states at the output of the curved part, the factor
$\sin\left[\pi\,(q^-_R - q^+_R)\right]$ appears in the output spin polarization,
which shows up as the oscillations in $P_\mathrm{out}(B)$ and $P_\mathrm{out}(R)$.

\begin{figure}
\includegraphics{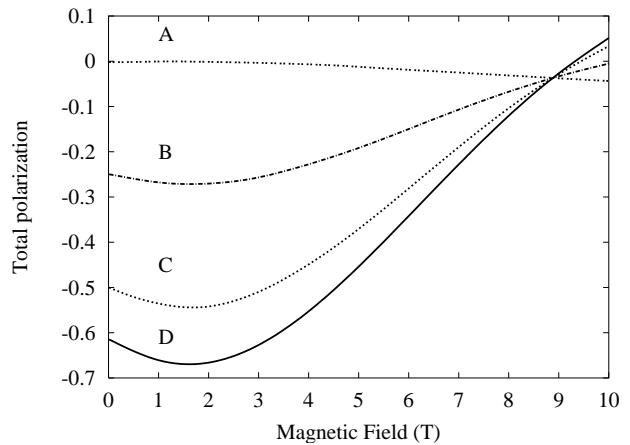}
\caption{\label{fig3}Total output polarization $P_\mathrm{out}$ vs. magnetic field for
different distribution width $w= \max\{\theta^{\pm}\}$ that
corresponds to the different degree of decoherence.
Curve (A) $w=2\pi$ (completely decoherent states), (B) $w=\pi$,
(C) $w=\pi/2$, (D) $w=0$ (completely coherent states).
Radius of the semi-circle is taken $5\cdot10^{-6}$ cm,
the other parameters are the same as in four previous figures. }
\end{figure}

Now we must say a few words about the influence of the initial phase difference
$\Delta\theta=\theta^- - \theta^+$ on the abovementioned effect. In general, the electron states
in the reservoirs are not coherent and, therefore, the output current densities
$j^{\pm}_{\mathrm{out}}$ have to be averaged over the distribution of random initial phases
$\theta^{\pm}$. In order to model the degree of decoherence we use rectangular distributions of
width $w$, $0\leq\theta^{\pm}\leq w$ for $\theta^{\pm}$.
The results are summarized in Fig.~\ref{fig3}. One can easily see, that initial
decoherence hampers the polarization. (A tiny polarization at strong magnetic fields
for completely decoherent case is due to the Zeeman effect only.)
Thus, these initial states must be specially prepared
in order to observe of the oscillating factor $\sin\left[\pi\,(q^-_R - q^+_R)\right]$
in the current density polarization. If the phase
difference $\theta^+ - \theta^-$ is not fixed in the electron beam, then
the observed polarization is always zero.

We would like to emphasize, that the current density redistribution
between the two spin-split modes in the curved part of the wire does not stem from the
finite curvature itself. 
In contrast, the {\em change} of the curvature
gives rise to the difference between 
the amplitudes $B^+$ and $B^-$ (or $D^+$ and $D^-$) and, therefore, leads to
the current density redistribution between the two spin-split modes.
Indeed, consider the electron momenta on the Fermi level for the straight and curved regions of the wire
in the simplest case of zero external magnetic field. The momenta
$k^{\pm}$ and $q^\pm$ are given by (\ref{Fermi_k_noB}) and (\ref{q_noB}) respectively.
The essential difference between the electron momenta for the straight wire and the loop lies in the
radius dependent term $\hbar^2/(2\alpha\,m^* R)$ in $q^\pm$.  Notice, that
$k^{\pm}=q^{\pm}/R$ if $R=\infty$ (no wire bending).
In contrast, $q^+/R$ {\em decreases} 
(as compared with $k^+$) and $q^-/R$ {\em increases}
(as compared with $k^-$) as long as the loop goes towards a kink of the wire $R\rightarrow 0$.
For all that,  the Fermi velocity $v_F=\hbar^2 k_0/m^*$ keeps the same
value in any part of the system, and
the electron momentum changes in curved part of the wire in such a way, that
{\em forward} scattering from one spin-split mode to another occurs. The latter leads to
the interference between them and shows up
as the current density redistribution between $j_\mathrm{out}^+$
and $j_\mathrm{out}^-$.
One can think about the wire bending as a changing of the initial 
Rashba parameter $\alpha$ to $\alpha\sqrt{1+\left[\hbar^2/(2\alpha\,m^* R)\right]^2}$
in the loop region.
Note, however, that in contrast to the actual change of $\alpha$, the change of
the wire's curvature does not affect the electron density of states and
results directly in the difference between $q^+$ and $q^-$ so, that 
there is no problem with the reflection.
This is a very particular property of the system: there is no barrier
on the junction between the straight and curved parts of the wire, but the
Fermi momenta have a jump, and, thus, the current density redistribution
between the spin-split modes takes place.

\subsection{$P_\mathrm{in}=1$}

In this section we propose to use a strongly curved 1D wire discussed above
as a spin switcher, i. e. the input current density
polarization  $P_\mathrm{in}$ is equal to $1$, and the output polarization can be 
switched to its opposite value.  Note, that in contrast to the previous case, the 
phase coherence of initial electron states is not necessary here.

\begin{figure}
\includegraphics{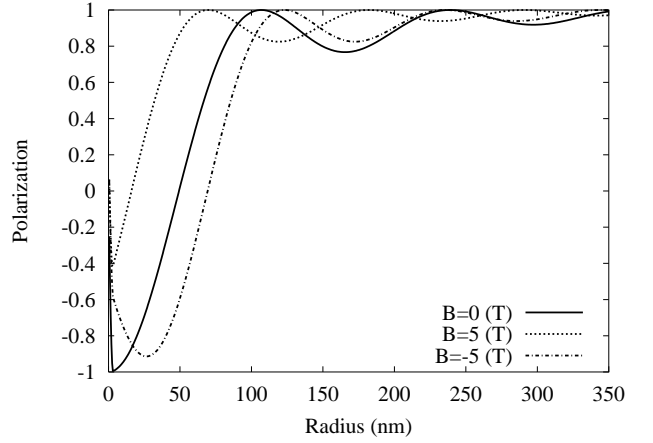}
\caption{\label{fig4} Polarization $P_\mathrm{out}$
versus radius of curvature at different external magnetic fields.
The input polarization $P_\mathrm{in}=1$, and
the other parameters are taken the same as in Fig.~\ref{fig2}.}
\end{figure}

The general solution of the system of equations (\ref{conditions})
at $A_0^+=1$, $A_0^-=0$
demonstrates zero backscattering ($R=0$ and $T=1$), while the polarization
curves exhibit the following interesting features (see Fig.~\ref{fig4}).
First, the plots of $P_\mathrm{out}(R)$ [as well as $P_\mathrm{out}(B_z)$) yield oscillating curves.
Second, the efficiency of the spin-switching depends strongly on the direction of the 
external magnetic field.
Third, although the polarization can be switched to its opposite value at $B_z=0$,
the relatively small radius of the wire's curvature is necessary.
In order to explain the features listed,
we solve the system (\ref{conditions}) in two cases again:
adiabatic  $\hbar^2/(2m^*R\alpha)\ll 1$ and strongly non-adiabatic $\hbar^2/(2m^*R\alpha)\gg 1$
limits.

The first limit is, however, not really interesting. As in the previous section,
no current density redistribution between the two spin-split modes occurs here, i. e.
$|D^+|^2=1$ and $|D^-|^2=0$. Intuitively it is clear, that the curved 
wire does not differ too much from the straight one as long as $\hbar^2/(2m^*R\alpha)\ll 1$.
Therefore, the polarization keeps its +100\% initial value while the current flows through the system.

In the opposite, strongly non-adiabatic limit, the situation changes 
drastically. Indeed, the system of equations (\ref{conditions})
at $\alpha^\pm=\pi/2$, $\beta^\pm=0$ and zero magnetic field allows
the following approximate solution 
\begin{eqnarray}
\nonumber &&
A^+=0, \quad A^- = 0, \\
\nonumber &&
B^+ = -\frac{i}{\sqrt{2}}{\mathrm e}^{i\,\theta^+ + \frac{i\,\pi}{2}\left(\frac{1}{2}+q^+_R\right)}, \\
\nonumber &&
B^- = -\frac{1}{\sqrt{2}}{\mathrm e}^{i\,\theta^+ + \frac{i\,\pi}{2}\left(q^-_R-\frac{1}{2}\right)},\\
\nonumber &&
C^+=0, \quad C^- = 0, \\
\nonumber &&
D^+ = \frac{1}{2}{\mathrm e}^{i\,\theta^+}
\left[{\mathrm e}^{i\,\pi\left(q^-_R - \frac{1}{2}\right)}
- {\mathrm e}^{i\,\pi\left(q^+_R + \frac{1}{2}\right)}\right],\\
\nonumber &&
D^- =\frac{1}{2\, i}{\mathrm e}^{i\,\theta^+}
\left[{\mathrm e}^{i\,\pi\left(q^-_R - \frac{1}{2}\right)}
+{\mathrm e}^{i\,\pi\left(q^+_R + \frac{1}{2}\right)}\right]. \\
\label{non-ad-solution_pol}
\end{eqnarray}
Then, 
$|D^\pm|^2=\frac{1}{2} \pm \frac{1}{2}\cos\left[\pi\,\left(q^+_R - q^-_R\right)\right]$,
and the spin components of the output current density read
\begin{equation}
j^\pm_\mathrm{out} = \frac{\hbar\, k_0}{2\,m^*}\left\{1 \pm \cos\left[\pi\left(q^-_R - q^+_R\right)\right]\right\}.
\end{equation}
Thus, the current density redistribution occurs in the strongly non-adiabatic regime, and
the polarization is
\begin{equation}
\label{p_estimation_pol}
P_\mathrm{out}=\cos\left[\pi\left(q^-_R - q^+_R\right)\right].
\end{equation}

Let us make some comments on (\ref{p_estimation_pol}). Note, that if
the radius of curvature is {\em exactly} equal to zero, then
the difference between the Fermi angular momenta reads 
\begin{equation}
\label{limR=0}
q^-_R - q^+_R=
\left.\frac{2m^*R\,\alpha}{\hbar^2}\sqrt{1+\left(\frac{\hbar^2}{2\alpha\,m^* R}\right)^2}\right|_{R=0}=1.
\end{equation}
Thus, the output polarization $P_\mathrm{out}=-1$, whereas the initial one was $P_\mathrm{in}=+1$.
Therefore, the polarization is switched to its opposite value at $R=0$ and $B_z=0$ as it is
explained in what follows.

Let us have an electron beam with $P_\mathrm{in}=1$ which is reflected
by an infinite barrier. If the dynamical and spin degrees of freedom are coupled
by means of the Rashba spin-orbit interactions then the reflected electron beam 
has exactly the opposite polarization $P_\mathrm{refl}=-1$.
Indeed, if the direction of the electron motion is perpendicular to the barrier then
a one-dimensional description is possible,
and the corresponding wave functions represent a sum of incident and reflected waves
[cf. with (\ref{input_psi+}) and (\ref{input_psi-})]
\begin{equation}
\label{psi_polar}
\psi^+(x)=\frac{1}{\sqrt{2}}{\mathrm e}^{i k^+ x}
\left(\begin{array}{c}
1 \\ -i \end{array} \right) + \frac{A^+}{\sqrt{2}}{\mathrm e}^{-i k^+ x}
\left(\begin{array}{c}
1 \\ i  \end{array} \right), 
$$
$$
\psi^-(x)=\frac{A^-}{\sqrt{2}}{\mathrm e}^{- i k^- x}
\left( \begin{array}{c}
i  \\
1 \end{array} \right).
\end{equation}
Then, we place an infinite barrier in the point $x=0$. Imposing
zero boundary conditions on (\ref{psi_polar}) we have the following system of equations
\begin{equation}
\label{spin-flip-eq}
\left\{\begin{array}{l}
1 + A^+ + iA^- = 0, \\ -i(1 - A^+) + A^- = 0. \end{array} \right.
\end{equation}
The solution of (\ref{spin-flip-eq}) is very simple and reads
\begin{equation}
\label{spin-flip-sol}
A^+ = 0, \qquad A^- = i.
\end{equation}
Substituting the amplitudes (\ref{spin-flip-sol}) into (\ref{psi_polar}),
we come to the conclusion, that due to the spin-orbit coupling 
the reflected electron beam always has the spin-polarization opposite to
the initial one.
Notice, that as soon as we assume zero radius of the curved part in Fig.~\ref{fig1}b we arrive 
at the one-dimensional wire with an infinite barrier. The finite 
radius and external magnetic field just gives some additional effects depicted in Fig.~\ref{fig4}.

The difference between the Fermi angular momenta depends not only on 
the Rashba coupling, but on the Zeeman splitting as well.
Therefore, the critical values of $q^-_R - q^+_R$, when the polarization $P$ changes the sign,
are tunable by means of the external magnetic field. Unfortunately, we do not have analytical formulae
for $q^\pm_{R,L}$ at non-zero magnetic fields, but one can see the effect in
Fig.~\ref{fig4}.
The most interesting curve is the one at $B_z=-5$T, where almost -100\% output
spin polarization is achieved at non-zero radius of the curvature.

\begin{figure}
\includegraphics{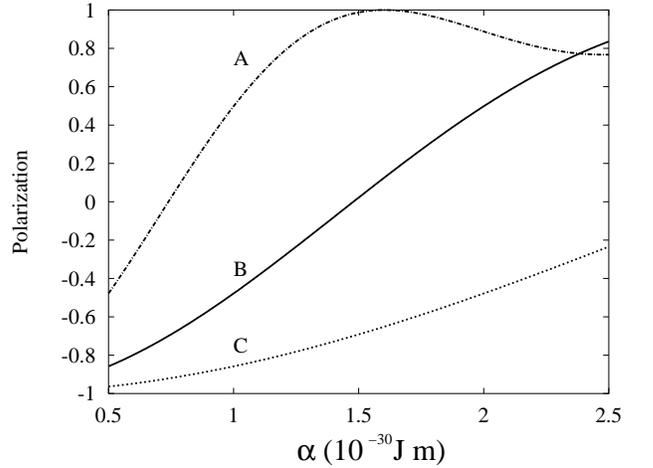}
\caption{\label{fig5} Polarization $P_\mathrm{out}$
versus Rashba constant $\alpha$ at zero external magnetic field.
The input polarization is $P_\mathrm{in}=1$, $m^*=0.033 m_e$, $E_F=30$meV, and
the radius of curvature is (A) $R=2\cdot 10^{-5}$cm, (B) $R=10^{-5}$cm, (c) $R=5\cdot 10^{-6}$cm.
Such values of $\alpha$ and $R$ are achievable experimentally in InAs
\cite{PRL2000grundler,PRB2000matsuyama,PRB2002yang}.}
\end{figure}

Such a spin-switcher can be  
utilized in devices of Datta-Das type \cite{APL1990datta,PRB2001mireles}
as reflectionless and high-speed spin-rotator.
To complete the spin field effect transistor we assume a spin polariser and 
a spin analyser at the ends of the input and output channels. 
For the sake of simplicity, let the spin polariser and spin analyser
be transparent for the same spin orientation.
The basic principle of the device proposed is similar to the ``conventional'' one.
The transistor is closed as long as the transport regime is adiabatic
$\hbar^2/(2\alpha m^* R) \leq 1$, and the electron spin
changes its orientation with respect to the spin-orientation in the contacts.
In contrast, the spin-switching occurs as soon as the electron spin
does not have enough time to follow the electron trajectory (non-adiabatic regime).
Thus, the spin-valve is opened when $\hbar^2/(2\alpha m^* R) \gg 1$.
The relation $\hbar^2/(2\alpha m^* R)$ can be tuned by the
gate-voltage dependent Rashba constant $\alpha$ as
it is discussed in Refs.~\cite{PRL2000grundler,PRB2000matsuyama}.
The plots of $P_\mathrm{out}(\alpha,P_\mathrm{in}=1)$ are shown in Fig.~\ref{fig5} for
different radii of curvature. The values of $\alpha$ are taken in accordance
with the experimental situation in InAs.\cite{PRL2000grundler,PRB2000matsuyama}

In contrast to the ``conventional'' spin-rotators made of the straight semiconductor stripes,
the device proposed is expected to operate faster since
it works in the non-adiabatic regime.
Indeed, the switching speed is determined 
by the time needed for an electron to propagate through the curved
part of the system which can be very short 
as long as our device is in the non-adiabatic regime $\hbar^2/(2\alpha m^* R) \gg 1$.
Thus, the switching time can be even smaller than the one
estimated in the Introduction for, let us say, a ``conventional''
spintronic device in the adiabatic regime.

\section{Conclusions}

In this paper, we investigated the ballistic transport in curved one-dimensional wires
with intrinsic spin-orbit interactions of Rashba type.
In detail, the projection of the current density spin-polarization
on the spin-quantization axis is considered.

The major points covered by this paper may be summarized as follows
(i) a strongly curved 1D wire with
Rashba spin-orbit coupling demonstrates
current density redistribution between the two spin-split modes without backscattering,
(ii) the current density redistribution shows up 
in the projection of its output spin-polarization 
on the spin-quantization axis defined by (\ref{polarization}),
(iii) strongly curved 1D wires with Rashba spin-orbit coupling
can switch the spin-polarization of the input electron beam
to the opposite one.

In our opinion, the main outcome of this paper is that
strongly curved 1D wires with Rashba spin-orbit coupling
can serve in the capacity of reflectionless and high-speed spin-switchers.
We believe that the interplay between Rashba spin-orbit coupling and
non-zero curvature of the one-dimensional system can find especially fruitful applications in spintronics.

\begin{acknowledgments}
The authors acknowledge financial support from DFG through
Graduiertenkolleg ``Physik nanostrukturierter Festk\"orper''  (M.T.)
and Sonderforschungsbereich 508 (A. C.).
\end{acknowledgments}

\appendix

\begin{widetext}
In Appendix, we adduce the system of equations solved above at zero external magnetic field
in two limiting cases. Deep in the adiabatic regime the equations (\ref{conditions}) read
\begin{eqnarray}
\nonumber &&
\left(A_0^+{\mathrm e}^{i\theta^+} + A^+ \right) - i \,\left(A_0^-{\mathrm e}^{i\theta^-}  - A^- \right)
= {\mathrm e}^{-i\,\pi /4} \left(B^+ {\mathrm e}^{-i\,\pi\, q_R^+ /2} + 
C^+  {\mathrm e}^{i \,\pi\, q_L^+ /2} -B^-  {\mathrm e}^{-i\,\pi\, q_R^- /2}
+ C^- {\mathrm e}^{i\,\pi\, q_L^- /2} \right), \\
\nonumber &&
\left(A_0^-{\mathrm e}^{i\theta^-} + A^- \right)   - 
 i \,\left(A_0^+{\mathrm e}^{i\theta^+} - A^+ \right)   
= {\mathrm e}^{-i\,\pi /4} \left(B^-{\mathrm e}^{-i\,\pi\, q_R^- /2}
 + C^-{\mathrm e}^{i\,\pi \, q_L^- /2} + B^+{\mathrm e}^{-i\,\pi \, q_R^+ /2} - 
C^+ {\mathrm e}^{i\,\pi \, q_L^+ /2}  \right), \\
\nonumber &&
{\mathrm e}^{-i\,\pi \,/4} \left(B^+{\mathrm e}^{i\,\pi \, q_R^+ /2} + 
C^+{\mathrm e}^{-i\,\pi \, q_L^+ /2} -  B^-{\mathrm e}^{i \,\pi \, q_R^- /2}+
C^-{\mathrm e}^{-i\,\pi \, q_L^- /2}\right) = 
D^+ + i \, D^-, \\
\nonumber &&
{\mathrm e}^{i\,\pi /4} \left(B^-{\mathrm e}^{i\,\pi\, q_R^- /2} + C^-{\mathrm e}^{-i\,\pi\, q_L^- /2}+ 
B^+{\mathrm e}^{i\,\pi\, q_R^+ /2} - C^+{\mathrm e}^{-i\,\pi \, q_L^+ /2} \right) = 
D^- + i \, D^+, \\
\nonumber &&
 k^+\left(A_0^+{\mathrm e}^{i\theta^+} -  A^+ \right) \,
 - i \, k^- \left(A_0^-{\mathrm e}^{i\theta^-}  +  A^- \right)
= \frac{{\mathrm e}^{i\,\pi /4}}{R} \left[
B^+\left(q_R^+ - \frac{1}{2}\right)\,{\mathrm e}^{-i\,\pi\, q_R^+ /2} -  \right. \\
\nonumber &&
\left. - C^+\left(\frac{1}{2} + q_L^+ \right) {\mathrm e}^{i\,\pi\, q_L^+ /2} - 
B^-\left( q_R^- - \frac{1}{2}\right){\mathrm e}^{-i\,\pi\, q_R^- /2} -
C^-\left(\frac{1}{2} + q_L^- \right){\mathrm e}^{i\,\pi\, q_L^- /2}
\right], \\
\nonumber &&
k^- \left(A_0^-{\mathrm e}^{i\theta^-} - A^- \right)   - i k^+ \,\left(A_0^+{\mathrm e}^{i\theta^+} +
A^+ \right)  = \frac{{\mathrm e}^{-i\,\pi /4 } }{R}\left[
B^-\left(\frac{1}{2} + q_R^- \right){\mathrm e}^{-i\,\pi\,q_R^- /2} + \right. \\
\nonumber &&
\left. +  C^-\left(\frac{1}{2} - q_L^- \right){\mathrm e}^{i\,\pi\, q_L^- /2}
+ B^+ \left(\frac{1}{2} + q_R^+ \right){\mathrm e}^{-i\,\pi \, q_R^+ /2} - 
C^+\left(\frac{1}{2} - q_L^+ \right){\mathrm e}^{i\,\pi \, q_L^+ /2} \right], \\
\nonumber &&
\frac{1}{R}{\mathrm e}^{-i\,\pi\, /4} \left[
B^+\left( q_R^+  - \frac{1}{2}\right){\mathrm e}^{i\,\pi\,q_R^+/2} -
C^+\left(\frac{1}{2} + q_L^+\right){\mathrm e}^{-i\,\pi\, q_L^+ /2} -
\right. \\
\nonumber &&
- \left. B^-\left( q_R^-  - \frac{1}{2}\right){\mathrm e}^{i\,\pi\,q_R^- /2}\, -
C^-\left( \frac{1}{2} + q_L^- \right){\mathrm e}^{-i\,\pi\,q_L^-/2}\right] = 
D^+\, k^+ +  i \,D^-\, k^- , \\
\nonumber &&
\frac{1}{R}{\mathrm e}^{i\,\pi /4} \left[
B^-\left(\frac{1}{2} + q_R^- \right){\mathrm e}^{i\,\pi\, q_R^- /2} + 
C^-\left(\frac{1}{2} - q_L^- \right){\mathrm e}^{-i\,\pi\, q_L^- /2}+ \right. \\
\nonumber &&
+ \left. B^+\left(\frac{1}{2} + q_R^+ \right){\mathrm e}^{i\,\pi\,q_R^+ /2} - 
C^+\left(\frac{1}{2} - q_L^+ \right) {\mathrm e}^{-i \,\pi q_L^+ /2} \right] = 
 D^-\, k^-  + i \,D^+ \, k^+ . \\
\label{system_adiabatic}
\end{eqnarray}
In the opposite, strongly non-adiabatic limit the system of equations can be written as
\begin{eqnarray}
\nonumber &&
\frac{1}{\sqrt{2}}\left(A_0^+{\mathrm e}^{i\theta^+} + A^+ \right) - 
\frac{i}{\sqrt{2}} \,\left( A_0^-{\mathrm e}^{i\theta^-}  - A^- \right)
= {\mathrm e}^{i\,\pi/4}
\left( C^+  {\mathrm e}^{i \,\pi\, q_L^+ /2} -B^-  {\mathrm e}^{-i\,\pi\, q_R^- /2} \right), \\
\nonumber &&
\frac{1}{\sqrt{2}}\left( A_0^-{\mathrm e}^{i\theta^-} + A^- \right)   - 
\frac{i}{\sqrt{2}} \,\left( A_0^+{\mathrm e}^{i\theta^+}  - A^+ \right)  
=  {\mathrm e}^{-i\,\pi/4}
\left(C^-{\mathrm e}^{i\,\pi \, q_L^- /2} + B^+{\mathrm e}^{-i\,\pi \, q_R^+ /2}\right), \\
\nonumber &&
{\mathrm e}^{-i\,\pi/4}\left(
C^+{\mathrm e}^{-i\,\pi \, q_L^+ /2} -  B^-{\mathrm e}^{i \,\pi \, q_R^- /2}\right) = 
\frac{1}{\sqrt{2}} D^+ + \frac{i}{\sqrt{2}} \, D^-, \\
\nonumber &&
{\mathrm e}^{i\,\pi/4}
\left( C^-{\mathrm e}^{-i\,\pi\, q_L^- /2} + B^+{\mathrm e}^{i\,\pi\, q_R^+ /2} \right)= 
\frac{1}{\sqrt{2}}D^- + \frac{i}{\sqrt{2}} \, D^+, \\
\nonumber &&
\frac{1}{\sqrt{2}}k^+\left(A_0^+{\mathrm e}^{i\theta^+} -  A^+ \right) 
 - \frac{i}{\sqrt{2}} \, k^- \left(A_0^-{\mathrm e}^{i\theta^-}  +  A^- \right)= \\
\nonumber &&
= \frac{{\mathrm e}^{i\,\pi /4}}{R} \left[
-C^+\left(\frac{1}{2} + q_L^+ \right) {\mathrm e}^{i\,\pi\, q_L^+ /2} -
B^-\left( q_R^- - \frac{1}{2}\right){\mathrm e}^{-i\,\pi\, q_R^- /2} \right], \\
\nonumber &&
\frac{1}{\sqrt{2}}k^- \left(A_0^-{\mathrm e}^{i\theta^-} - A^- \right)   - 
\frac{i}{\sqrt{2}} k^+ \,\left( A_0^+{\mathrm e}^{i\theta^+} + A^+ \right) = \\
\nonumber &&
= \frac{{\mathrm e}^{-i\,\pi /4 } }{R}\left[ 
C^-\left(\frac{1}{2} - q_L^- \right){\mathrm e}^{i\,\pi\, q_L^- /2} 
+ B^+ \left(\frac{1}{2} + q_R^+ \right){\mathrm e}^{-i\,\pi \, q_R^+ /2} \right], \\
\nonumber &&
\frac{1}{R}{\mathrm e}^{-i\,\pi\, /4} \left[
- C^+\left(\frac{1}{2} + q_L^+\right){\mathrm e}^{-i\,\pi\, q_L^+ /2} -
B^-\left( q_R^-  - \frac{1}{2}\right){\mathrm e}^{i\,\pi\,q_R^- /2}\right] = 
\frac{1}{\sqrt{2}} D^+\, k^+  +   \frac{i}{\sqrt{2}} \,D^-\, k^- , \\
\nonumber &&
\frac{1}{R}{\mathrm e}^{i\,\pi /4} \left[
C^-\left(\frac{1}{2} - q_L^- \right){\mathrm e}^{-i\,\pi\, q_L^- /2}+ 
B^+\left(\frac{1}{2} + q_R^+ \right){\mathrm e}^{i\,\pi\,q_R^+ /2} \right] = 
\frac{1}{\sqrt{2}} D^-\, k^-  + \frac{i}{\sqrt{2}} \,D^+ \, k^+ . \\
\label{system_non-adiabatic}
\end{eqnarray}
\end{widetext}

\bibliography{disser.bib}

\end{document}